\pdfoutput=1

\documentclass[]{aastex631}

\usepackage{xcolor, soul}

\def\gtorder{\mathrel{\raise.3ex\hbox{$>$}\mkern-14mu
             \lower0.6ex\hbox{$\sim$}}}
\def\ltorder{\mathrel{\raise.3ex\hbox{$<$}\mkern-14mu
             \lower0.6ex\hbox{$\sim$}}}

\begin{document}

\title{SINGLE OBJECT OBSERVATIONS: LARGE TELESCOPE VS. MULTIPLE SMALL TELESCOPES}

\author{S. Ben-Ami}
\affiliation{Department of particle physics and astrophysics, Weizmann Institute of Science, 76100 Rehovot, Israel.}

\correspondingauthor{S. Ben-Ami}
\email{sagi.ben-ami@weizmann.ac.il}

\author{E.~O.~Ofek}
\affiliation{Department of particle physics and astrophysics, Weizmann Institute of Science, 76100 Rehovot, Israel.}

\begin{abstract}
A significant fraction of large-telescope observing time is devoted to single-object spectroscopy and photometry. We compare the cost-effectiveness of different telescope architectures, including single seeing- and diffraction-limited telescopes and arrays of seeing-limited or diffraction-limited telescopes. We use the survey speed per unit cost as the performance metric. For single telescopes with equal collecting area, diffraction-limited operation becomes more cost-effective once seeing-limited observations enter the background-dominated regime, typically at visible magnitudes of  $\sim20$. For unresolved single-object observations, arrays of seeing-limited telescopes are generally more cost-effective than a single seeing-limited telescope. When matched in survey speed, a single diffraction-limited telescope outperforms an array of small seeing-limited telescopes only for targets fainter than $\sim20$--$24\,$mag, while a single large diffraction-limited telescope becomes more cost-effective than an array of smaller diffraction-limited telescopes only for targets fainter than $\sim26$--$28\,$mag. We conclude with a discussion on aspects related to the minimum telescope size and maintenance of telescope arrays. These conclusions are sensitive to the specific assumptions and system parameters adopted for each architecture and should therefore be interpreted with appropriate caution. Our results motivate the development of telescope arrays such as MAST and LFAST and may inform the design of future optical facilities. 
\end{abstract}

\keywords{
instrumentation: miscellaneous ---
methods: observational ---
telescopes}

\section{Introduction}
\label{sec:Introduction}

A large fraction of observing time on large telescopes is dedicated for single-object observations. This mainly includes seeing-dominated spectroscopy and imaging. For example,  the Low Resolution and Imaging Spectrograph (LRIS; \citealt{Oke+1995_LRIS}) on the Keck I telescope continues to be used for an average of more than 100 nights per year (\citealt{Fucik2024LRIS2}). The construction of additional larger telescopes is limited by the cost and complexity of such projects. Therefore, when constructing telescope systems, we would like to maximize their cost-effectiveness.
The first step is to define a figure of merit that will allow us to analyze the problem. However, this figure of merit depends on our goals. For example, in \cite{Ofek+BenAmi2020_Grasp_SkySurvrys_CostEffectivness} we analyzed the cost-effectiveness of a survey telescope.
Suppose that one defines the main goal of a survey telescope as maximizing the number of targets the telescope can observe per unit of time. In that case, a reasonable figure of merit is the grasp, defined as the relative volume of space a telescope can observe, to some $S/N$, per unit of time.
In the background-noise-dominated regime (i.e., sources fainter than magnitude 19-21, depending on sky background at the site), the grasp is given by
\begin{equation}
    \mathcal{G} \propto \Omega \Big(\frac{S}{N}\Big)^{-3/2} A_{\rm eff}^{3/4} B^{-3/4} \sigma^{-3/2} \frac{t_{\rm E}^{3/4}}{t_{\rm E} + t_{\rm D}  }.
    \label{eq:grasp_basic}
\end{equation}
Here, $\Omega$ is the telescope field of view (solid angle on sky), $A_{\rm eff}$ is the effective collecting area, $B$ is the sky background in photons per unit area, per unit time, per unit solid angle on sky, $\sigma$ is the image quality, $t_{\rm E}$ is the exposure time, {and $t_{\rm D}$ is the dead time - see Table \ref{tab:parameters} for a list of parameters appearing in this publication}. Our analysis suggests that, for the background-noise-dominated regime, once the detector pixel drops below $\sim5\,\mu$m, multiple small telescopes are more cost-effective than a single seeing-limited large telescope. Such small pixels became available only around the year 2020, and led to the construction of cost-effective survey telescopes like the Large Array Survey Telescope (LAST; \citealt{Ofek+2023PASP_LAST_Overview, BenAmi+2023PASP_LAST_Science}),
and the planned Argus array (\citealt{Law+2022PASP_ArgusArray}).
This statement is correct only for seeing-limited telescopes, whereas for a diffraction-limited survey telescope $\mathcal{G}\propto A_{\rm eff}^{3/2}$, and therefore larger telescopes are more cost-effective, compared to multiple small telescopes. A detailed discussion of the grasp and its limitations is presented in \cite{Ofek+BenAmi2020_Grasp_SkySurvrys_CostEffectivness}.

For the case of single-object imaging or spectroscopy, a different figure of merit is required. The figure of merit we choose to use is the survey speed, defined as the number of single sources, with flux $f$, that can be observed to a predefined signal-to-noise ratio $\left(S/N\right)$ per unit of time. Since narrow-field-of-view observations can be obtained using adaptive optics, we consider the two branches of single-object observations: seeing-dominated and diffraction-limited. Here we derive the survey speed expression in the relevant regimes and use it to draw several conclusions related to the cost-effectiveness of different kinds of telescope systems.

The analysis summarized in this work provides the rationale for the construction of telescope arrays for single-object spectroscopy and photometry. In recent years, several such efforts were initiated. These include: The Multi-Aperture Spectroscopic Telescope (MAST; Ben-Ami et al. in prep.; \citealt{Irani+2024SPIE_MAST_DeepSpec}; \citealt{SoferRimlat+2024arXiv_MAST_HighSpec}), LFAST (\citealt{Angel+2022SPIE_LFAST_SpectroscopyTelescope,Bender2024}), PolyOculus (\citealt{Eikenberry+2019_PolyOculus}), and the Pan-chromatic Array for Survey Telescopes (PAST; Garrappa et al., in prep.).

In \S\ref{sec:SN} we discuss the $S/N$, and in \S\ref{sec:ReadNoise} we analyze the impact of the detector read-noise. In \S\ref{sec:speed} we derive the formulae for the survey speed. The cost-effectiveness is discussed in \S\ref{sec:CostEff}, and we conclude in \S \ref{sec:Discussion}.

\section{The signal-to-noise}
\label{sec:SN}
We begin by deriving the well-known signal-to-noise ratio expressions for several generic observation scenarios involving {\it unresolved} single sources. As shown below, the results depend on the transition between the source-noise-dominated and background-noise-dominated regimes; accordingly, we employ the full signal-to-noise expression rather than the limiting cases appropriate to a single dominant noise source. We neglect detector readout noise throughout this analysis, a simplification that is justified in \S \ref{sec:ReadNoise}.  In observations, one typically differentiates between $S/N$ for measurement processes (e.g., photometry, morphology, or astrometry) and $S/N$ for detection processes (e.g., \citealt{Zackay+2017_CoadditionI}). 
Throughout the paper, we only treat the case of single target observations (spectroscopy or photometry with accuracy worse than a few millimagnitudes). Given this motivation, we are mainly interested in the $S/N$ for measurements.

The $S/N$ for a measurement process is:
\begin{equation}
    (S/N)_{\rm meas} = \frac{\epsilon\alpha_{\rm f}\mathcal{R}^{-1}f A_{\rm eff} t_{\rm E}}{\sqrt{A_{\rm eff} t_{\rm E}\mathcal{R}^{-1}( {\alpha_{B}\sigma^2 }B +\epsilon\alpha_ff ) }}.
    \label{eq:SNR}
\end{equation}
Here $f$ is the source photon flux in the desired band (e.g., units of photons\,cm$^{-2}$\,s$^{-1}$), $A_{\rm eff}$ is the effective collecting area in the desired band (e.g., units of cm$^{2}$), $t_{\rm E}$ is the exposure time (e.g., units of seconds), $B$ is the background in band (e.g., units of photons\,cm$^{-2}$\,s$^{-1}$\,arcsec$^{-2}$). $\epsilon$ is the Strehl ratio ($\cong1$ for seeing-limited observations). $\alpha_{\rm f}$  is the fraction of source flux captured (e.g., in the slit), ${\alpha_{\rm B}\sigma^2}$ is the effective area over which background noise is collected - see further information in the appendix, and $\sigma$ is the standard deviation of the PSF (e.g., $FWHM/2.35$ for a Gaussian PSF). {We provide some approximations for $\alpha_{f}$ and $\alpha_{B}$ in Appendix~\ref{app:alpha}}.
We also introduce $\mathcal{R}$, the approximate spectral resolution within the effective band. The latter scaling will be useful for evaluating the dependence of survey speed on spectral resolution.

In the background-noise dominated case, or $S/N$ for a detection process, the $\epsilon \alpha_{f} f$ term in the denominator should be set to zero. In the case of a source-noise dominated case, $B$ should be set to zero.
In the case of diffraction-limited observations, we use the relation between the Airy disk and the aperture area (for simplicity, we write the following equations in specific units):
\begin{equation}
\sigma_{\rm arcsec}^{2}\cong\eta\lambda_{0.6}^2/A_{{\rm d, cm}^{2}},    
\label{eq:sigma2A2}
\end{equation}
where $\lambda_{0.6}$ is the wavelength in $0.6\,\mu$m (we assume a circular aperture\footnote{For discussion of non-circular aperture, see e.g., \cite{Nir+2019_NoncircularPupilTelescope}.}), $A_{d,{\rm cm}^{2}}$ is the effective area in cm$^{2}$ (subscript $\rm d$ marks the diffraction-limit case), $\eta$ is a constant, that for our default units is $\cong23.04$,
and $\sigma_{\rm arcsec}$ is the PSF standard deviation in arcsec\footnote{The FWHM of Airy disk is $\cong1.028\lambda/D$, where $D$ is the telescope diameter. To be consistent with the Gaussian PSF, here we define $\sigma$ of a diffraction-limited telescope to be $(1.028/2.35)\lambda/D$.}.
Substituting into Equation \ref{eq:SNR}, we derive the following expression for the $S/N$ of a diffraction-limited telescope:
\begin{equation}
    ({S}/{N})_{\rm diff}
=
\frac{
\epsilon \alpha_{\rm f} f (\mathcal{R}^{-1} A_{\rm d} t_{\rm E})^{1/2}
}{
\sqrt{
\alpha_{\rm B} B 
\frac{23 \lambda_{0.6}^{2}}{A_{{\rm d, cm}^{2}}}
+ \epsilon \alpha_{\rm f} f
}
}.
\end{equation}
The transition from source noise to background noise takes place when the variance of the source noise equals the variance of the background noise. Writing this equality and converting to magnitudes, we get:
\begin{equation}
    m_{S\leftrightarrow B} = m_{\rm B} + 2.5\log_{10}(\alpha_{\rm f}/\alpha_{\rm B})-5log_{10}\sigma,
    \label{eq:Mag_Back2Src}
\end{equation}
where $m_{\rm B}$ is the sky background magnitude [mag\,arcsec$^{-2}$].
For resolved sources, $\sigma$ is the source size.
We find that under 1-arcsec seeing and low sky background levels, the transition occurs at magnitudes 19--22. For diffraction-limited telescopes, the transition shifts to fainter magnitudes as the aperture grows, and so the PSF occupies a progressively smaller area on the sky.
For diffraction-limited observations, the transition magnitude, $m_{S\leftrightarrow B}$ is pushed towards fainter magnitudes. Therefore, a comparison between diffraction-limited and seeing-limited telescopes should include the full $S/N$ formula (Equation~\ref{eq:SNR}) and not only the background/source-noise dominated approximations.

\section{Read noise}
\label{sec:ReadNoise}

So far, we have ignored the system readout-noise ($R$). Since in the read-noise dominated regime $S/N\propto t_{\rm E}$, rather than $S/N\propto t_{\rm E}^{1/2}$, when we design a system we attempt, if possible, to avoid the read-noise dominated regime. The transition between the read-noise-dominated regime and the background-noise-dominated regime takes place at an exposure time of (\citealt{Ofek+2024AJ_DetectionStrategy_OortCloud_Quadrature}):
\begin{equation}
    t_{R\leftrightarrow B} \approx \frac{R^{2}}{p^{2} A_{{\rm cm}^{2}} 10^{-0.4(m_{\rm B}-14.76)} }\,{\rm s},
\end{equation}
where $p$ is the pixel size in arcsec\,pix$^{-1}$.
For $R=2$\,e$^{-}$, $p=1$\,arcsec\,pix$^{-1}$, and $m_{B}=21$\,mag\,arcsec$^{-2}$, this results in observations being readout noise dominated for exposures of $4$\,s, $0.2$\,s, and $0.006$\,s or shorter for 20\,cm, 1\,m, and $5\,$m telescopes respectively.

To estimate the exposure time for the read-noise-dominated regime for spectroscopy, we compare:
\begin{equation}
Bt/\mathcal{R} = R^{2}/(p_{\lambda}p),
\end{equation}
which gives:
\begin{equation}
t_{R\leftrightarrow B} \approx \frac{R^{2}\mathcal{R}}{p_{\lambda}p A_{{\rm cm}^{2}} 10^{-0.4(m_{\rm b}-14.76)} }\,{\rm s},
\end{equation}
where $p_{\lambda}$ is the pixel scale per spectral resolution element.

We see that readout noise can play an important role for low-resolution spectrographs on small telescopes, or high-resolution spectrographs for all telescopes. While this seems like a significant limitation for small telescopes, several solutions exist, including low readout-noise detectors such as EM-CCD, qCMOS, and qCCD.
In the case of the Multi Aperture Spectroscopic Telescope (MAST; Ben-Ami et al. in prep), the low-resolution spectrograph DeepSpec ($\mathcal{R}\approx650$) uses standard CCDs, and the system becomes readout-noise dominated for exposure times lower than $\mathcal{O}(10^2)\,$s (\citealt{Irani+2024SPIE_MAST_DeepSpec}). In the case of MAST HighSpec ($\mathcal{R}\approx20,000$), we take advantage of the low readout-noise of an EM-CCD (\citealt{SoferRimalt2024}), and so the transition occurs for similar exposure times, despite the higher spectral resolution. In the following discussion, we will ignore the readout noise and assume it is being mitigated in the specific system design.

\section{Survey speed}
\label{sec:speed}

\begin{figure}
{
\centerline{    \includegraphics[width=9.2cm]{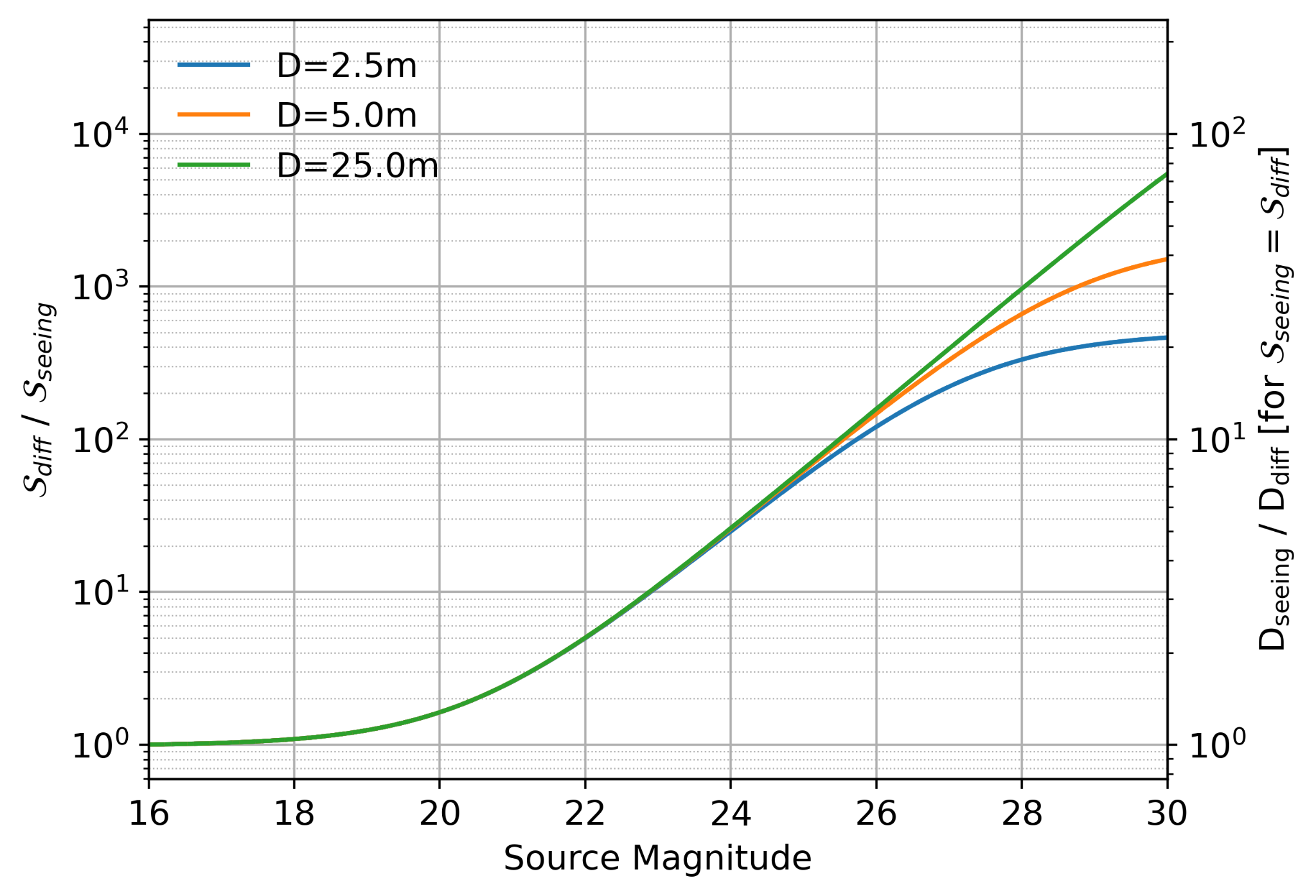}}}
\caption{Survey-speed ratio between diffraction-limited and seeing-limited telescopes with the same aperture
as a function of the target $V$-band magnitude.
The right Y-axis shows the aperture ratio required for a seeing-limited telescope to match the survey speed of a diffraction-limited telescope (Equation~\ref{eq:Ds}).
Assumed representative parameters for all plots: the sky background, from which $B$ is calculated, is $m_{\rm B}=21.6$\,mag\,arcsec$^{-2}$; $\epsilon_{s}=\epsilon_{d}=1$; $\eta=23.04$; $\alpha_{\rm f}=1$; $\alpha_{\rm B}=14$; We assume that a $V$-band source with AB magnitude 0 produces $10^{6}$\,photons\,cm$^{-2}$\,s$^{-1}$.
\label{fig:SurveySpeed}}
\end{figure}

While for a survey telescope, in the background-noise dominated regime, the effectiveness is given by the grasp (\citealt{Ofek+BenAmi2020_Grasp_SkySurvrys_CostEffectivness}; i.e., volume per unit time), for a single target (e.g., narrow field), a better way to describe the effectiveness of a telescope system is by its survey speed -- the number of targets with a given flux that can be observed at a specified $S/N$ by the telescope in a unit of time. For single-target observations we assume that $t_{\rm D}\ll t$.
The survey speed is given by:
\begin{equation}
    \mathcal{S}\propto t_{E}^{-1} .
\end{equation}
To calculate the survey speed, we isolate $t_{E}^{-1}$ from the $S/N$ (Equation~\ref{eq:SNR}).
The survey speed for the seeing-limited case is:
\begin{equation}
   \mathcal{S}_{\rm seeing} = \frac{\ {A_s}\,{{\epsilon_{\rm s}}}^2\,{{\alpha_{\rm f}}}^2\,f^2}{\mathcal{R}{(S/N)}^2 \left(B\,{\alpha_{\rm B}}\,\sigma^2+{\epsilon_{\rm s}}\,{\alpha_{\rm f}}\,f\right)}.
\end{equation}
Here, we replace $\epsilon$ with $\epsilon_{\rm s}$ which is $\cong1$, and replace $A$ by $A_{\rm s}$, to distinguish it from the diffraction-limited case. 
Next, substituting Equation~\ref{eq:sigma2A2}, the survey speed for the diffraction-limited case is:
\begin{equation}
    \mathcal{S}_{\rm diff} = \frac{{{A_d}}^2\,{{\epsilon_{\rm d}}}^2\,{{\alpha_{\rm f}}}^2\,f^2}{\mathcal{R}{(S/N)}^2\left(\eta\,B\,{\alpha_{\rm B}}\,{\lambda _{0.6}}^2+{A_d}\,{\epsilon_{\rm d}}\,{\alpha_{\rm f}}\,f\right)}.
\end{equation}
Here, $\epsilon_{\rm d}$ is the Strehl ratio of the diffraction-limited telescope,
and we replaced $A$ by $A_{d}$ (i.e., the collecting area of a diffraction-limited telescope).
The ratio between the survey speeds is:
\begin{equation}
    \frac{\mathcal{S}_{\rm diff}}{\mathcal{S}_{\rm seeing}} = \frac{{{A_d}}^2 {{\epsilon_{\rm d}}}^2 \left(B {\alpha_{\rm B}} \sigma_{\rm s} ^2+{\epsilon_{\rm s}} {\alpha_{\rm f}} f\right)}{{A_s} {{\epsilon_{\rm s}}}^2 \left(\eta B\,{\alpha_{\rm B}} {\lambda _{0.6}}^2+{A_d} {\epsilon_{\rm d}} {\alpha_{\rm f}} f\right)}
\end{equation}
In Figure~\ref{fig:SurveySpeed}, we present the survey speed ratio as a function of the source magnitude. 
This is plotted assuming $A_{d}=A_{s}$, $\epsilon_{d}=\epsilon_{s}=1$, and sky background of 21.6\,mag\,arcsec$^{-2}$. The other parameters are listed in the caption of Figure~\ref{fig:SurveySpeed}.
The first break in the plot around magnitude 20 is due to the transition from source-noise dominated to background-noise dominated in seeing-limited telescopes, while the second transition (and separation of telescope diameters) around magnitude 26 is similar but for diffraction-limited telescopes of different diameters.

\section{Cost Effectiveness}
\label{sec:CostEff}
\cite{vanBelle+2004_TelescopeCostScaling}  found that telescope cost (\$) scales like \$$\propto A^{1.3}$ when surveying telescopes with primary mirror diameters of $1-12\,$m. 
While this relation shows some scatter, it provides an adequate basis for comparing different telescope system design approaches. Using these relations, we define the telescope survey speed cost effectiveness ($\mathcal{C}^{\mathcal{S}}\propto\mathcal{S}/\$$). Here, we calculate the cost-effectiveness ratio between different types of approaches for telescope construction (e.g., diffraction-limited vs. seeing-limited and a single telescope vs. an array of telescopes). However, this discussion is limited to specific science cases: specifically, single-target spectroscopy and  photometry at a precision level larger than a few millimagnitudes.
These results do~not hold for other science cases like high-resolution imaging,
precision photometry and astrometry, and multi-object spectroscopy.
The reasons the following results are not necessarily relevant for these specific science cases are that,
some of them require the diffraction limit of a large telescope,
or the $S/N$ of these science cases may contain additional terms (e.g., scintillation noise for ground-based photometry and astrometry; e.g.,  \citealt{Young1967AJ_Photometry_Scintialtions_ReigerTheoryConfirmation, Lindegren1980_AtmosphericScintilation_GroundBasedAstrometry, Ofek2019_Astrometry_Code}),
or that the instrument may dominate the cost. These caveats should be borne in mind to avoid overinterpretation of the results.

We find the following expression for cost effectiveness for seeing limited observation and diffraction limit observations:
\begin{equation}
    \mathcal{C}^{\mathcal{S}}_{\rm seeing} \propto \frac{ {A_s^{-0.3}}\,{{\epsilon_{\rm s}}}^2\,{{\alpha_{\rm f}}}^2\,f^2}{\mathcal{R}{(S/N)^2 \left(B\,{\alpha_{\rm B}}\,\sigma_{\rm s} ^2+{\epsilon_{\rm s}}\,{\alpha_{\rm f}}\,f\right)}},
\end{equation}
and
\begin{equation}
    \mathcal{C}^{\mathcal{S}}_{\rm diff} \propto \frac{{{A_d^{0.7}}}\,{{\epsilon_{\rm d}}}^2\,{{\alpha_{\rm f}}}^2\,f^2}{\mathcal{R}{(S/N)}^2\left(\eta\,B\,{\alpha_{\rm B}}\,{\lambda _{0.6}}^2+{A_d}\,{\epsilon_{\rm d}}\,{\alpha_{\rm f}}\,f\right)}.
    \label{eq:C_diff}
\end{equation}

The cost-effective ratio between a {\it single} seeing-limited telescope with aperture area of $A_s$ and a {\it single} diffraction-limited telescope with aperture area $A_d$ is:
\begin{equation}
    \frac{\mathcal{C}^{\mathcal{S}}_{diff}}{\mathcal{C}^{\mathcal{S}}_{seeing}} = 
     \frac{{{A_d^{0.7}}}{A_s^{0.3}}\,{{\epsilon_{\rm d}}}^2\,\left(B\,{\alpha_{\rm B}}\,\sigma_{\rm s} ^2+{\epsilon_{\rm s}}\,{\alpha_{\rm f}}\,f\right)}{{{\epsilon_{\rm s}}}^2\,\left(\eta\,B\,{\alpha_{\rm B}}\,{\lambda _{0.6}}^2+{A_d}\,{\epsilon_{\rm d}}\,{\alpha_{\rm f}}\,f\right)}
\end{equation}
For telescopes of identical collecting power (i.e., $A_d=A_s$), the cost-effectiveness ratio is reduced to the survey speed relation shown in Figure \ref{fig:SurveySpeed}.

Since the cost function of a telescope grows faster than its collecting area (\citealt{vanBelle+2004_TelescopeCostScaling}), it is clear that an array of small telescopes may have some advantages (e.g., \citealt{Ofek+BenAmi2020_Grasp_SkySurvrys_CostEffectivness}). This is reinforced when considering that for small-aperture telescopes, the power law of the cost scaling relation is even steeper. For example, the cost of $60\,\mathrm{cm}$-aperture versus $100\,\mathrm{cm}$-aperture systems from leading vendors in the market follows a power-law index of approximately $1.6$. 

We turn to derive the cost-effectiveness ratio between a single (large) seeing-limited telescope and an array of $N_{\rm tel}$ seeing-limited (small) telescopes. Assuming the array has the same collecting area as the single large telescope and hence the same survey speed, the cost-effectiveness ratio is:
\begin{equation}
     \frac{\mathcal{C}^{\mathcal{S}}_{{\rm seeing, a}}}{\mathcal{C}^{\mathcal{S}}_{{\rm seeing, 1}}} \propto N_{\rm tel}^{-1}\left( \frac{A_{\rm s,1}}{A_{\rm s,a}} \right)^{1.3}= \left( \frac{A_{\rm s,1}}{A_{\rm s,a}} \right)^{0.3},
    \label{eq:arraySrc}
\end{equation}
where subscript '$\mathrm{a}$'  marks a telescope in the array, while subscript '$1$' marks the single large aperture telescope. $N_{\rm tel}$, the number of telescopes in the array, is equal to the area ratio between the single-aperture large telescope and a telescope in the array.

\begin{figure}
    \centering
    {
    \includegraphics[width=0.48\textwidth]{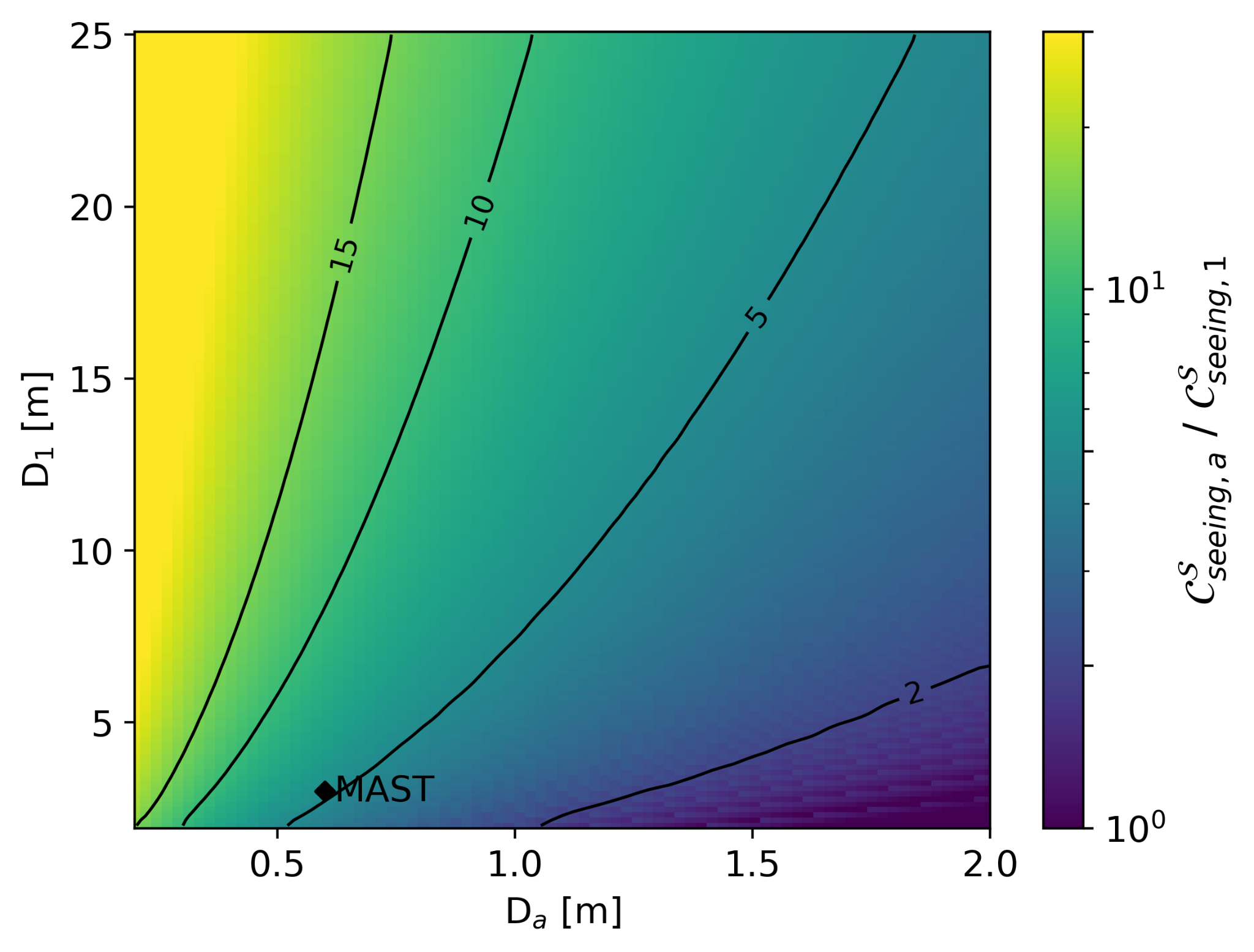}}
    \caption{Cost-effectiveness ratio for a single-aperture telescope versus a telescope array for seeing-limited observations. The two systems are required to have equivalent survey speeds. We include a steepening of the power law-index to 1.6 for aperture diameters $\leq2\,\mathrm{m}$. Parameters used are listed in Figure \ref{fig:SurveySpeed} caption.
   }
    \label{fig:CostRatio_arraySrc}
\end{figure}

This relation, shown in Figure \ref{fig:CostRatio_arraySrc}, is straightforward and implies that, for seeing-limited system, an array of small telescopes is often more cost-effective compared to a single large telescope when considering, e.g., single-object spectroscopy.
An additional advantage of such arrays is the ability to operate sub-arrays, which increases overall system flexibility.

We now turn to estimate the cost-effectiveness ratio between a diffraction-limited telescope and a seeing-limited array. We begin by deriving the relationship between a seeing and diffraction-limited telescope diameters that yields equal survey speed:  
\begin{equation}
D_{\mathrm{seeing}}^{=\mathcal{S}} =
\frac{\sqrt{\pi} D_{\mathrm{diff}}^{2}\,\epsilon_{d}}{\epsilon_{s}}
\sqrt{
\frac{
B\mathcal{R} \alpha_{\mathrm{B}} \sigma^{2}
+
\epsilon_{s}\alpha_{\mathrm{f}} f
}{
4\eta B \alpha_{\mathrm{B}} \lambda_{0.6}^{2}
+
\pi D_{\mathrm{diff}}^{2} \epsilon_{d}\alpha_{\mathrm{f}} f
}
},
\label{eq:Ds}
\end{equation}

\begin{figure*}
    \centering
   
    \begin{minipage}{0.48\textwidth}
        \centering
        \includegraphics[width=\linewidth]{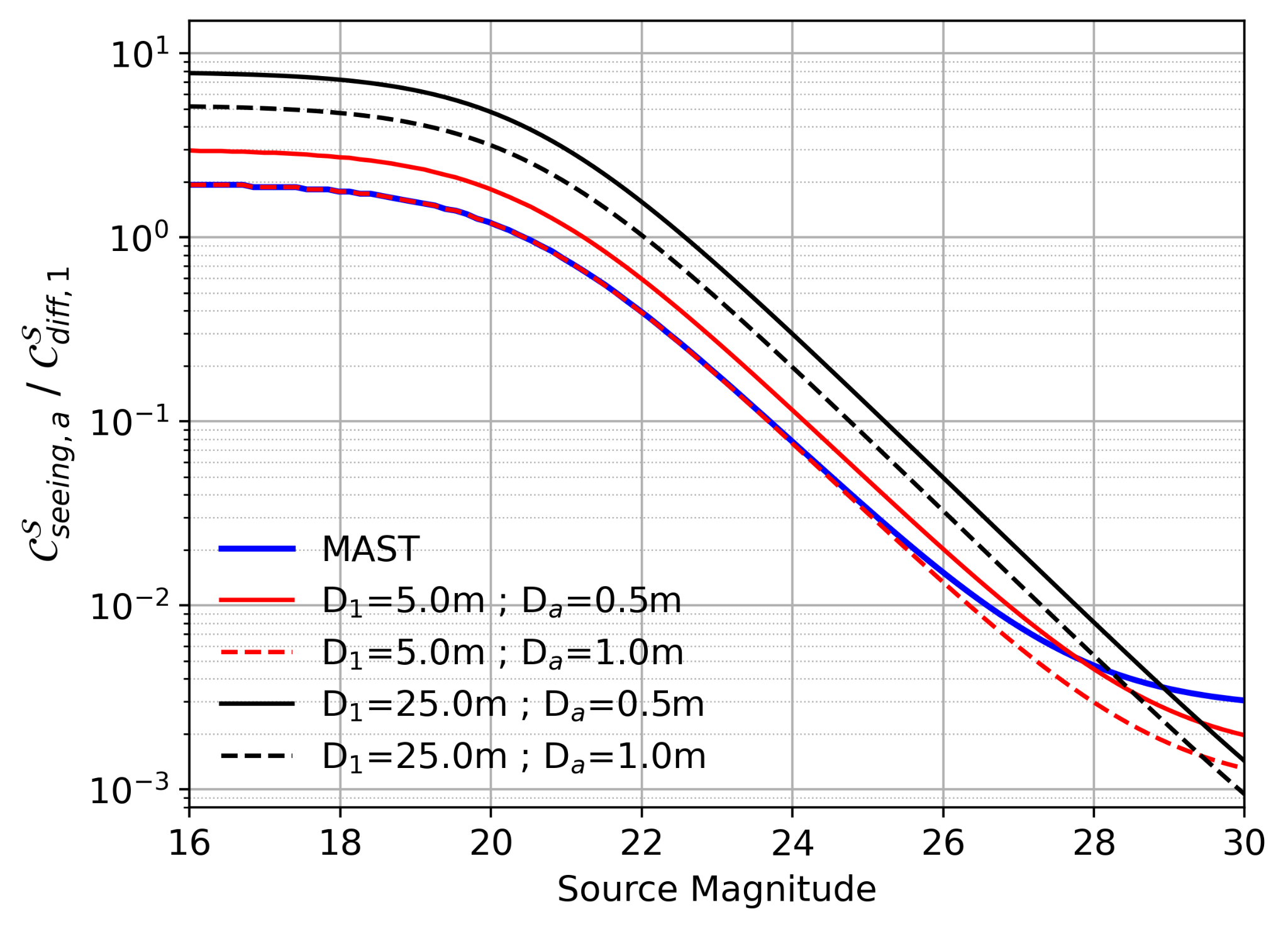}
    \end{minipage}
    \hfill
    \begin{minipage}{0.48\textwidth}
        \centering
        \includegraphics[width=\linewidth]{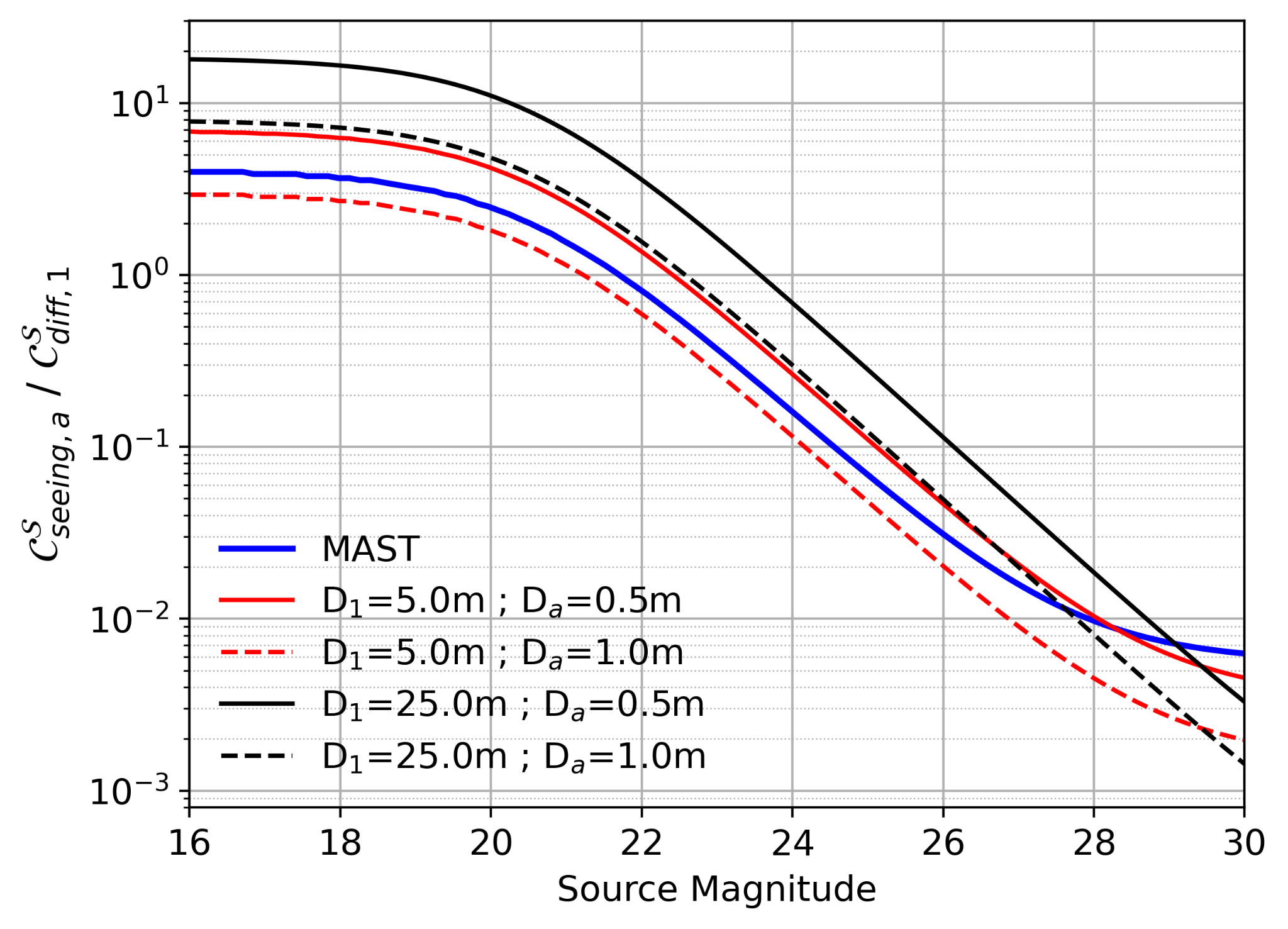}
    \end{minipage}

    \caption{The cost-effectiveness ratio between an array of seeing-limited telescopes and a single diffraction-limited telescope with the same survey speed as the array. The cost effectiveness is plotted as a function of the source V-band magnitude.  Left: Telescope costs are assumed to follow the scaling relation of \cite{vanBelle+2004_TelescopeCostScaling}; Right: We include a steepening of the power law-index to 1.6 for aperture diameters $\leq2\,\mathrm{m}$. Parameters used are listed in Figure \ref{fig:SurveySpeed} caption.}
    \label{fig:slArray_vs_dl}
\end{figure*}

Requiring the array and the single telescope to achieve equal survey speed, the number of telescopes in the array should be (using Equation~\ref{eq:Ds}):
\begin{equation}
N_{\rm tel}\approx
\left(
\frac{D_{\mathrm{seeing}}^{=\mathcal{S}}}{D}
\right)^{2}
=
\frac{A_{\rm diff}^{2}}{A_{\rm eff}}
\left(\frac{\epsilon_d}{\epsilon_s}\right)^2
\,
\frac{
B \alpha_{\mathrm{B}} \sigma^{2}
+
\epsilon_{s}\alpha_{\mathrm{f}} f
}{
\eta B \alpha_{\mathrm{B}} \lambda_{0.6}^{2}
+
A_{\rm diff}\,\epsilon_{d}\alpha_{\mathrm{f}} f
}.
\end{equation}
and we obtain the following cost-effectiveness ratio: 
\begin{equation}
  \frac{\mathcal{C}^{\mathcal{S}}_{seeing, \rm a}}{\mathcal{C}^{\mathcal{S}}_{diff, \rm 1}} 
=
\frac{
\epsilon_s^{2}
\left( \eta\,B\,\alpha_{\rm B}\,\lambda_{0.6}^{2}
+ A_{d,1}\,\epsilon_d\,\alpha_{\rm f}\,f \right)
}{
A_{d,1}^{0.7} A_{s,a}^{0.3}\,\epsilon_d^{2}
\left( B\,\alpha_{\rm B}\,\sigma_s^{2}
+ \epsilon_s\,\alpha_{\rm f}\,f \right)
},
\label{eq:CseeingA_Cdiff1}
\end{equation}
Figure~\ref{fig:slArray_vs_dl} shows Equation~\ref{eq:CseeingA_Cdiff1} for some representative visible light photometric or spectroscopic observations.
As in Figure~\ref{fig:SurveySpeed}, the first break around magnitude 20 is due to the transition from the source-noise to background-noise dominated regimes, in the seeing-dominated telescope. The second break, around magnitude 28, is due to the same transition in the diffraction-limited telescope. For larger telescopes, this transition shifts to fainter magnitude, and therefore it is not seen in the case of $D_{diff,1}=25$\,m.

When the aperture ratio between the diffraction-limited and the seeing-limited telescope increases, the cost-effectiveness of the seeing-limited system increases.
Assuming that fully effective diffraction-limited telescopes are a few times more expensive than seeing-limited telescopes with the same size, the plot suggests that diffraction-limited telescopes become more cost-effective for targets fainter than magnitude 22 to 24.

\begin{figure*}
    \centering
{    
    \begin{minipage}{0.48\textwidth}
        \centering
        \includegraphics[width=\linewidth]{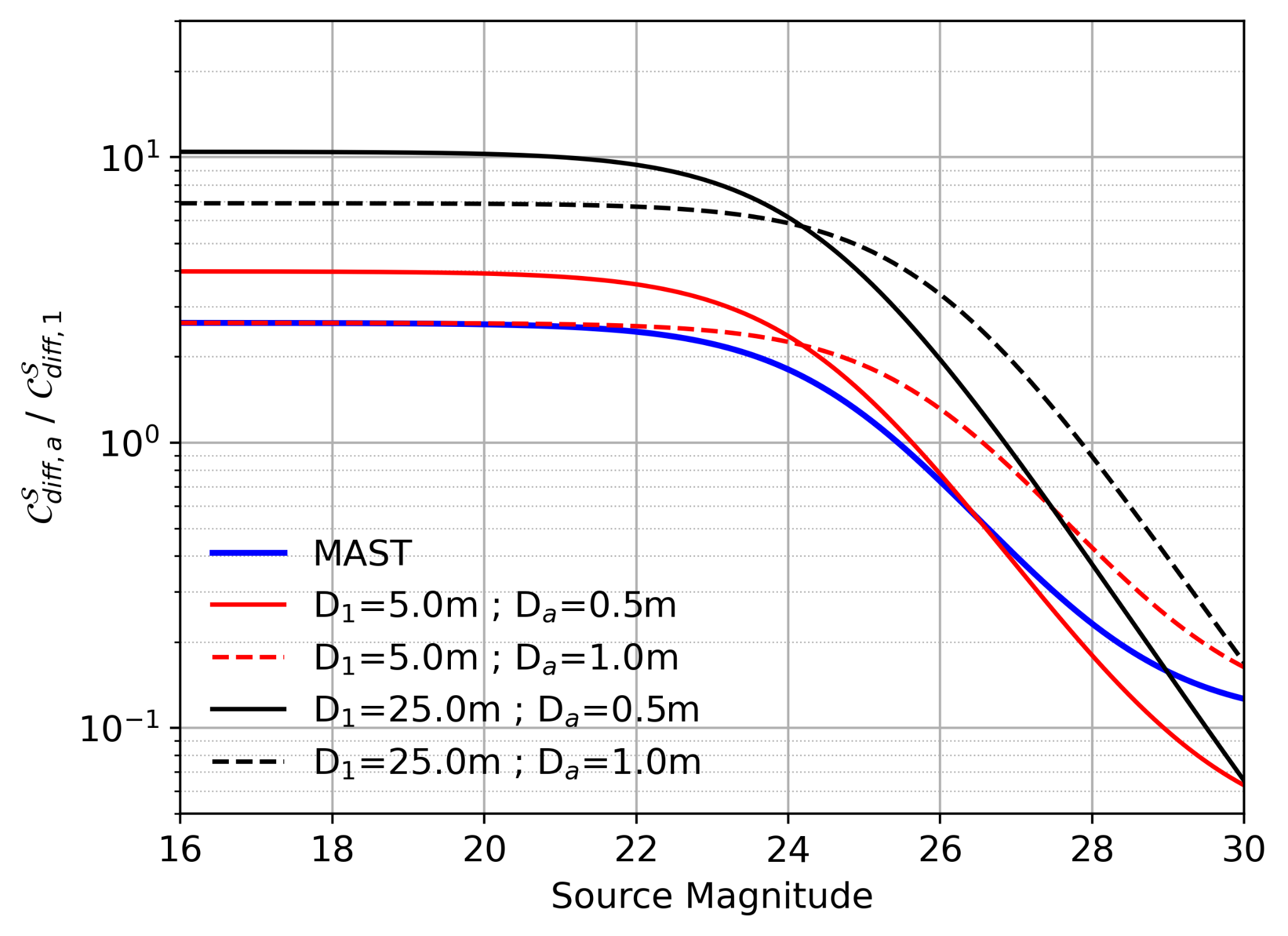}
    \end{minipage}
    \hfill
    \begin{minipage}{0.48\textwidth}
        \centering
        \includegraphics[width=\linewidth]{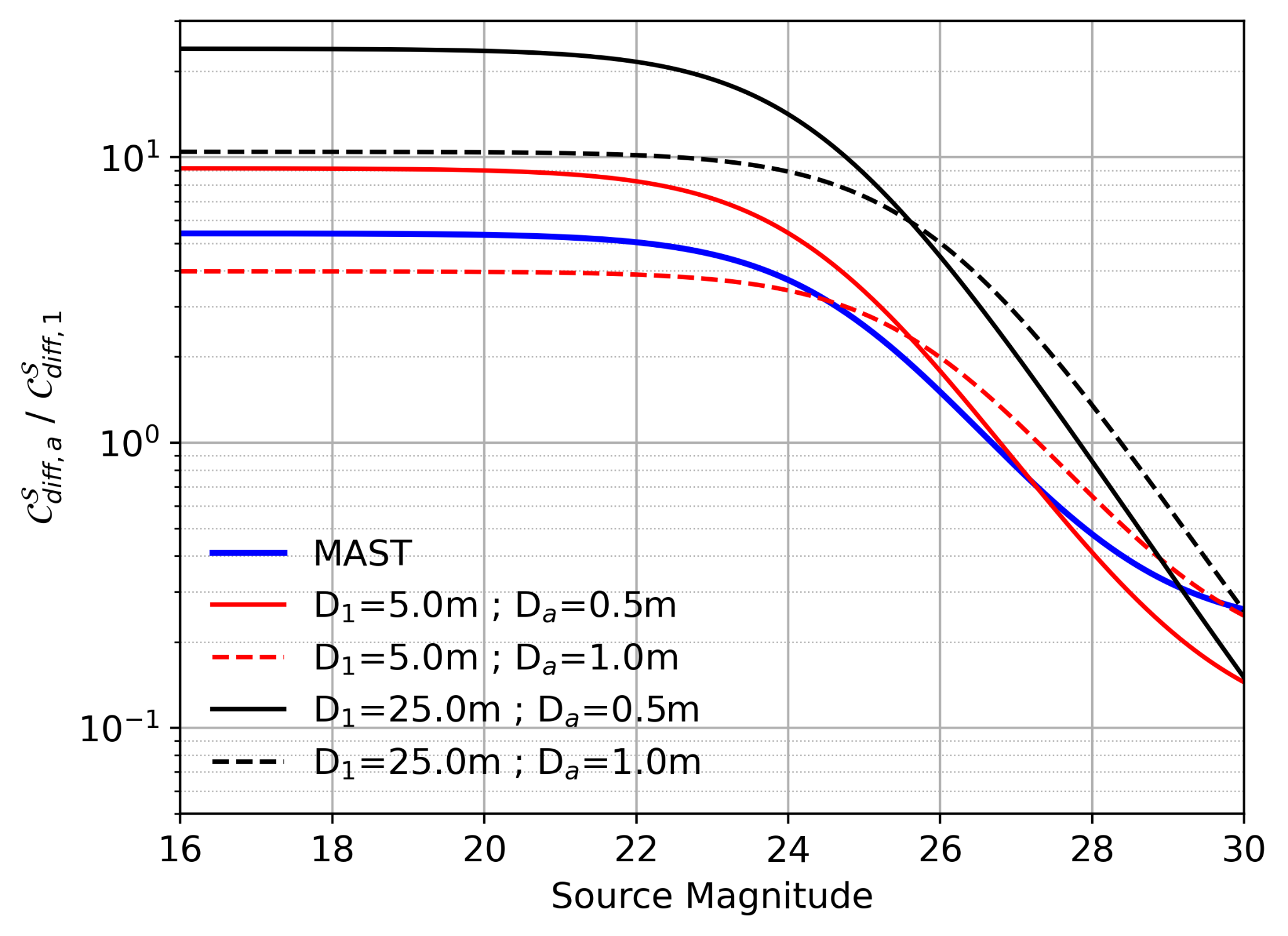}
    \end{minipage}

    \vspace{1em}

}
    \caption{
Cost-effectiveness ratio of a diffraction-limited telescope array (element diameter $D_{\rm a}$) relative to a diffraction-limited single telescope of diameter $D_{\rm 1}$ at equal survey speed. Left: Telescope costs are assumed to follow the scaling relation of \cite{vanBelle+2004_TelescopeCostScaling}; Right: We include a steepening of the power law-index to 1.6 for aperture diameters $\leq2\,\mathrm{m}$. Parameters used are listed in Figure \ref{fig:SurveySpeed} caption.}
    \label{fig:dlArray_vs_dl}
\end{figure*}

The final case we investigate is when both the array and the single telescope are operating in the diffraction limit. As before, equal collecting area does~not provide equal survey speed, so we first derive the number of small diffraction-limited telescopes required to achieve the same survey speed as a single large diffraction-limited telescope:

\begin{equation}
 N_{\rm tel}
    =
    \frac{A_{d,1}^2}{A_{d,a}^2}
    \frac{
    \eta\,B\,\alpha_{\rm B}\,\lambda_{0.6}^2
    + A_{d,a}\,\epsilon_{\rm d}\,\alpha_{\rm f}\,f
    }{
    \eta\,B\,\alpha_{\rm B}\,\lambda_{0.6}^2
    + A_{d,1}\,\epsilon_{\rm d}\,\alpha_{\rm f}\,f
    }.
\end{equation}
Using this equation, we can now calculate the cost-effectiveness ratio between an array of diffraction-limited (small) telescopes and a single diffraction-limited telescope, assuming equal survey speed:
\begin{equation}
\frac{\mathcal{C}^{\mathcal{S}}_{diff, \rm a}}{\mathcal{C}^{\mathcal{S}}_{diff, 1}} 
=
\left(\frac{A_{d,a}}{A_{d,1}}\right)^{0.7}
\frac{
\eta\,B\,\alpha_{\rm B}\,\lambda_{0.6}^{2.}
+ A_{d,1}\,\epsilon_{\rm d1}\,\alpha_{\rm f}\,f
}{
\eta\,B\,\alpha_{\rm B}\,\lambda_{0.6}^{2}
+ A_{d,a}\,\epsilon_{\rm da}\,\alpha_{\rm f}\,f}
\end{equation}
Results are shown in Figure \ref{fig:dlArray_vs_dl} and suggest that an array of diffraction-limited telescopes may be more cost-effective for targets brighter than about 26 to 28 magnitude. We conclude that incorporating adaptive optics systems to small telescopes will push the transition in which arrays are no longer cost-effective to fainter magnitudes. 

We emphasize three assumptions in our analysis that may affect the quantitative results, though not the qualitative conclusions:
(i) We assume that the diffraction-limited telescope delivers perfect wavefront correction, $\epsilon_d=1$, which is not achievable with current adaptive optics (AO) systems, particularly at short wavelengths ($\lambda \lesssim 650\,\mathrm{nm}$ ; but see e.g., \citealt{Males2024MagAOX});
(ii) We do not include the cost of the AO system (or launch to space) required to maintain diffraction-limited performance across the relevant magnitude range. However, this cost ratio is likely roughly a multiplication factor to Equation~\ref{eq:C_diff}. Based on a few, not necessarily representative examples,
this factor is of the order of $2$--$4$ for ground-based telescopes and about $100$ for large space-based telescopes;
and (iii) we do not account for the efficiency of AO system, both in terms of throughput and dead time (e.g., settling time), which would further reduce survey speed. However, this is likely another small multiplication factor ($\lesssim 2$).

 \section{Discussion}
 \label{sec:Discussion}

The analysis presented here applies to spectroscopic and photometric observations of single, {\it unresolved} sources. Other science cases may impose different requirements, involve different cost functions, and follow different signal-to-noise formulations. For example, small telescope arrays are unlikely to compete with a single large telescope in applications such as multi-object spectroscopy or high-resolution diffraction-limited observations. {In these regimes, despite the higher cost of instrumentation and increased system complexity—such as larger optical elements, more demanding optical designs associated with wider fields of view, and the challenges of manufacturing slit masks or deployable multi-fiber systems—a single large telescope is likely to remain the more cost-effective solution. However, in the context of current transient-survey follow-up, it is worth noting that even at high alert rates, the mean surface density of follow-up targets is typically less than one per field of regard for most telescopes. Consequently, multi-object spectroscopy is unlikely to offer significant gains for rapid transient characterization.}

Nevertheless, single-object spectroscopy and photometry currently account for a substantial fraction of the observing time on large telescopes. In this context, the present analysis is well motivated and highlights distinct performance and cost-effectiveness behaviors for seeing-limited and diffraction-limited facilities. These differences should be carefully considered when planning the next generation of optical observatories.    
Furthermore, this analysis assumes that we would like to shift from an era in which one telescope can do "everything" to an era of specialized facilities. We now proceed to summarize the results derived in the previous sections and discuss their broader implications.

 The survey speed for measurement process of seeing-limited and diffraction-limited telescopes exhibits different dependencies on aperture, with the governing behavior set by the transition between the source-noise and background-noise regimes. In the source-noise limit, the dominant noise contribution originates from the intrinsic Poisson fluctuations of the source, and so diffraction-limited and seeing-limited telescopes of similar diameters exhibit similar survey speeds. However, once the seeing-limited system operates in the background-limited regime, the performance diverges. 

When these survey-speed trends are combined with empirical telescope cost relations, with $\mathcal{C} \propto A^{1.3}$ for a single-aperture facility, a contrasting pattern of cost effectiveness emerges. Under seeing-limited conditions, the comparatively weak scaling of survey speed with aperture causes the survey-speed-to-cost ratio to decline as telescopes become larger. In the diffraction-limited regime, by contrast, the strong scaling of survey speed with aperture improves the cost-effectiveness of larger systems. These distinctions form the basis for the cost-effectiveness results presented in this work and delineate the observational regimes in which diffraction-limited telescopes offer a clear performance advantage over their seeing-limited counterparts.

We also find that telescope arrays composed of modest-aperture elements ($\lesssim 1\,$m) can, in some regimes, deliver significantly higher survey speed per unit cost than single-aperture telescopes. This advantage arises because the survey speed of a seeing-limited system, and of diffraction-limited telescopes operating in the source noise limit, increases more slowly than the cost. We argue that for smaller aperture telescopes, this is more profound, as the power law of the cost scaling relation is even steeper. For example, the cost of $60\,\mathrm{cm}$-aperture versus $100\,\mathrm{cm}$-aperture systems from leading vendors in the market follows a power-law index of approximately $1.6$.

Furthermore, when comparing a seeing-limited array to a diffraction-limited telescope of equal survey speed, the former remains more cost-effective down to sources roughly one to two magnitudes fainter than the sky background. These findings help justify the emergence of next-generation array-based facilities, such as MAST \citep[Ben-Ami et al., in prep; ][]{Irani+2024SPIE_MAST_DeepSpec, SoferRimlat+2024arXiv_MAST_HighSpec}, LFAST \citep{Angel+2022SPIE_LFAST_SpectroscopyTelescope, Bender2024}, PolyOculus \citep{Eikenberry+2019_PolyOculus}, and PAST (Garrappa et al. in prep.). We conclude that as the demand for efficient single-object follow-up continues to grow, telescope arrays are poised to play an increasingly central role in meeting both the scientific and operational challenges of modern observational astronomy.

For designing array elements, practical constraints impose a natural range of apertures. In imaging applications, the smallest useful diameter is set by the need to avoid the diffraction limit dominating over atmospheric seeing, implying $D\gtrsim 20$\,cm. For fiber-fed single-object spectroscopy, fiber-core diameters and focal ratio degradation requirements impose further constraints. Assuming we want to avoid using single-mode fibers and guarantee high-efficiency coupling without an AO system, the on-sky angular size of a few-mode fiber with a $\sim20\,\mu$m core is:
\begin{equation}
    \phi \cong 1.65 \frac{D_f}{20\,\mu{\rm m}}
\left(\frac{D}{0.5\,{\rm m}}\right)^{-1}
\left(\frac{F}{4}\right)^{-1}\,{\rm arcsec},
\end{equation}
which implies that for typical $1-2$\,arcsec seeing, apertures of $\gtrsim0.5$\,m are needed to maintain high coupling efficiency (we assume beam convergence should be in the range of F/3 to F/5). Coupled with cost-scaling considerations, this identifies a practical design range of approximately $0.5$--$1.5$\,m for array elements. This can be bypassed by using a relay system to speed up the beam at the focal plane, at the cost of additional optical elements for each telescope. In any realistic system analysis, losses associated with fiber coupling and attenuation must be accounted for; such losses are largely absent in, e.g., long-slit spectroscopy with large single-aperture telescopes. 

Operational considerations further support the feasibility of telescope arrays. Although the maintenance burden naively increases with the number of telescopes, small units offer substantial practical benefits: they are easier to access, require minimal heavy equipment, and can often be serviced by non-specialist personnel.
The experience of LAST, which currently operates 40 telescopes totaling 1.8\,m of effective aperture with only $\sim0.25$--$0.5$ full-time employees on-site, illustrates the practicality of telescope arrays.

The trends identified here demonstrate that cost-effective performance does not necessarily require a single large aperture, but rather the appropriate alignment of aperture, image quality, and system architecture. As these considerations become increasingly central to future observatory planning, array-based designs are poised to play a defining role in the next generation of optical facilities.

\begin{acknowledgments}
We thank  Nick Konidaris, Chad Bender, Shri Kulkarni, Avishay Gal-Yam, and Barak Zackay for their valuable discussions.

S.B.A. is grateful for support from the André Deloro Institute for Advanced Research in Space and Optics, Peter and Patricia Gruber Award, Willner Family Leadership Institute for the Weizmann Institute of Science, Aryeh and Ido Dissentshik Career Development Chair, and Israel Ministry of Science (grant 3-18140).

E.O.O. is grateful for support by grants from the Willner Family Leadership Institute, Madame Olga Klein-Astrachan, André Deloro Institute, Schwartz/Reisman Collaborative Science Program, Paul and Tina Gardner, The Norman E Alexander Family Foundation ULTRASAT Data Center Fund, Jonathan Beare, Israel Science Foundation, Vatat, Minerva, and BSF.
\end{acknowledgments}

\begin{table}
\centering
\caption{Summary of parameters.}
\label{tab:parameters}
{
\begin{tabular}{|l|l|}
\hline
\textbf{Parameter} & \textbf{Description} \\
\hline

$\mathcal{G}$ & Grasp (survey volume per unit time). \\
$\Omega$ & Telescope field of view (solid angle on sky). \\
$S/N$ & Signal-to-noise ratio. \\
$A_{\rm eff}$ & Effective collecting area (cm$^{2}$). \\
$A_s, A_d$ & Collecting area for seeing-limited and diffraction-limited telescopes. \\
$A_{\rm 1}, A_{\rm a}$ & Collecting area of single telescope and array element. \\

$B$ & Sky background flux (photons cm$^{-2}$ s$^{-1}$ arcsec$^{-2}$). \\
$m_{\rm B}$ & Sky background magnitude (mag arcsec$^{-2}$). \\

$\sigma$ & Standard deviation of the PSF (arcsec). \\
$\sigma_s$ & PSF width in the seeing-limited case. \\

$t_{\rm E}$ & Exposure time (s). \\
$t_{\rm D}$ & Dead time between exposures (s). \\

$f$ & Source photon flux (photons cm$^{-2}$ s$^{-1}$). \\
$n_{\nu}$ & Photon flux density. \\
$r$ & Photon detection rate. \\

$\epsilon, \epsilon_s, \epsilon_d$ & Strehl ratio (general, seeing-limited, diffraction-limited). \\

$\alpha_{\rm f}$ & Fraction of source flux captured (e.g., slit or aperture throughput). \\
$\alpha_{\rm B}$ & Factor defining effective background collection area. \\

$\mathcal{R}$ & Spectral resolution. \\
$p$ & Pixel scale (arcsec pixel$^{-1}$). \\
$p_{\lambda}$ & Pixel scale per spectral resolution element. \\

$N_{\mathrm{pix}}$ & Number of independent spectral channels (pixels). \\
$N_{\mathrm{tel}}$ & Number of telescopes in the array. \\

$\lambda_{0.6}$ & Wavelength normalized to $0.6\,\mu$m. \\
$\eta$ & Constant relating PSF size to aperture in diffraction limit. \\

$R$ & Detector readout noise (electrons). \\

$t_{R\leftrightarrow B}$ & Exposure time at transition between read-noise and background-noise regimes. \\
$m_{S\leftrightarrow B}$ & Magnitude at transition between source-noise and background-noise regimes. \\

$\mathcal{S}$ & Survey speed (number of targets per unit time). \\

$\mathcal{C}^{\mathcal{S}}$ & Cost-effectiveness (survey speed per unit cost). \\

$D$ & Telescope diameter. \\

$d$ & Slit width. \\
$h$ & Effective height of slit trace. \\
$r_{\rm f}$ & PSF fitting radius. \\
$r_{\rm a}$ & Aperture radius. \\

\hline
\end{tabular}}
\end{table}

\appendix
\section{Definitions of aperture factors}
\label{app:alpha}

In the equations above ${\alpha_{\rm f}}$ and ${\alpha_{\rm B}\sigma^2}$  depend on the operation conditions and the type of observation (e.g., PSF photometry, aperture photometry).

Assuming a circularly symmetric Gaussian PSF, for PSF-weighted photometry we get:
\begin{eqnarray}
    \alpha_{\rm f} \approx [1 - \exp(-r_{\rm f}^{2}/(2\sigma^{2})]\cong1,\\
    {\alpha_{\rm B}\sigma^2} \approx \frac{\pi FWHM^2}{2\ln(2)}\cong12.5\sigma^2,
    \label{eq:alphaPSF}
\end{eqnarray}
where $r_{\rm f}$ is the fit radius of the PSF. For aperture photometry:
\begin{eqnarray}
    \alpha_{\rm f} \approx [1 - \exp(-r_{\rm a}^{2}/(2\sigma^{2})]\cong 1,\\
    {\alpha_{\rm B}\sigma^2} \approx \pi r_a^2\cong6\pi\sigma^2,
    \label{eq:alphaAper}
\end{eqnarray}
where $r_{\rm a}$ is the aperture radius, which we assume in the last step is equal to a point source image FWHM. In practice, the aperture size may depend on other factors. Finally, for slit photometry and spectroscopy (\textit{i.e.,} the light fraction in a slit with width $d$, fitted with a 1-D Gaussian PSF of width $\sigma$):
\begin{eqnarray}
    \alpha_{\rm f} \approx {\rm erf}\Big(\frac{d}{2\sqrt{2}\sigma}\Big)\cong 1,\\
    {\alpha_{\rm B}\sigma^2} \approx d\times h\cong14\sigma^2,
    \label{eq:alphaSlit}
\end{eqnarray}
where erf is the error function, and the slit width is taken to be slightly larger than a point source FWHM, and we assume the 2-D trace is collapsed into a 1-D trace of a height of $2\times FWHM$. At this stage, we ignore the readout noise, which will be discussed in \S\ref{sec:ReadNoise}.

\bibliographystyle{aasjournal}
\bibliography{papers,papers2}

\end{document}